\documentstyle[12pt,aasms4]{article}

\slugcomment{Submitted to ApJ Letters}

\lefthead{Kuijpers}
\righthead{A Solar Prominence Model}

\begin{document}

\newcommand{\grad}{\vec{\nabla}}
\newcommand{\rot}{\vec{\nabla}\times}
\newcommand{\dov}{\vec{\nabla}\cdot}
\newcommand{\A}{\vec A}
\newcommand{\B}{\vec B}
\newcommand{\be}{\begin{equation}}
\newcommand{\beq}{\begin{eqnarray}}
\newcommand{\e}{\epsilon_0}
\newcommand{\ee}{\end{equation}}
\newcommand{\eeq}{\end{eqnarray}}
\newcommand{\E}{\vec E}
\newcommand{\ep}{\epsilon}
\newcommand{\f}{\frac}
\newcommand{\m}{\mu_0}
\newcommand{\mb}{\vec m}
\newcommand{\me}{\frac{1}{4\pi\epsilon_0c^2}}
\newcommand{\mm}{\frac{\m}{4\pi}}

\title{A Solar Prominence Model}

\author{Jan~Kuijpers}
\affil{
Sterrekundig Instituut, Universiteit Utrecht,
Postbus 80\,000,   3508 TA Utrecht\\
  Dep. of Exp. High Energy Physics, University of
Nijmegen, 6525~ED Nijmegen\\
  CHEAF, 1009~DB Amsterdam, The Netherlands
}

\begin{abstract} We propose a model for solar prominences based on
converging flow observed in the chromosphere and photosphere. In contrast
with existing models we do not apply a shearing motion along the neutral
line. Instead we assume that bipolar loops approaching on different sides
of the neutral line have a non-vanishing magnetic helicity of the {\it
same} sign. In the converging flow the individual loops kink and develop a
skew. For loops of the same helicity the skew is in the same sense. As a
result the `chiral' symmetry of an aligned distribution of loops is broken
and the reconnecting loop system forms a filament with the observed
magnetic orientation and anchoring of the barbs in regions of parasytic
polarity. The filament consists of a number of individual strands of
coaxial coronal electric currents each of which is current neutralised.
The filament material is suspended in dips in the magnetic field and the
transverse field direction coincides with that in the Kuperus-Raadu model.
Above the filament a cavity forms with an overlying arcade consisting of
the outer portions of the reconnected loops.  \end{abstract}

\keywords{Sun: magnetic fields --- Sun:  filaments --- Sun:
prominences --- Sun: corona --- MHD}

\section{Introduction}~\label{intro}

Solar prominences or filaments occur above boundaries of {\it converging}
photospheric flows (\cite{mart86,romp86}), located between opposite
polarity magnetic fields (\cite{marti90}). Along these `polarity
inversions' magnetic flux disappears (flux cancelling). 

The intermediate legs or appendages along the sides of a quiescent
filament are rooted in magnetic fields opposite in polarity to the network
magnetic fields on the same side (\cite{martetal94,martech94}). Not only
is the perpendicular component of the field reversed but also the axial
component is opposite to what would be expected from differential rotation
acting on coronal arcades (\cite{leroy83}). This `inverse' configuration
has been called the `Double Inverse Polarity Paradigm' (\cite{kup96}). 

Quiescent filaments are predominantly {\it dextral} -- i.e.  the field
direction in the filament is to the right for an imaginary observer in the
chromosphere on the positive-polarity side and facing the broad side of
the filament -- in the {\it northern} solar hemisphere, and sinistral in
the southern hemisphere, independent of the solar cycle
(\cite{martetal94}). 

Below I will take these observations -- the {\it converging migration of
fields,} and the association of the ends of a filament with network
magnetic fields of {\it opposite polarity} -- as the starting point for an
evolutionary model of a quiescent filament. I will show that a successful
filament model obtains if one starts off with bipolar loops with the {\it
same magnetic helicity} on both sides of the polarity inversion line. No
assumption on the existence of a shearing flow will be needed. The current
structure of the proposed filament model consists of an aligned row of
`neutralised' currents, each with coaxial current closure.  The mass and
the mass motions inside the filament result from intermittent internal and
external reconnections, subsequent impulsive heating, chromospheric
evaporation, cooling and condensation inside individual twisted flux
tubes.

\section{Filament formation from reconnecting\\ 
twisted loops} 

\subsection{A single flux tube in converging flow}

Consider a single twisted flux tube, rooted below the photosphere. In the
ideal MHD approximation the free magnetic energy of the tube -- stored in
coronal electric currents -- {\it increases} as the footpoints approach
each other. This can be seen as follows: the free magnetic energy of a
tube $W_f$ is given by the azimuthal component of the magnetic field,
which derives from the axial current. As the number of windings between
the footpoints of an individual field line is fixed in ideal MHD while the
tube length $L$ decreases, the azimuthal field component varies
approximately as $B_\phi = B_z R/L$, and increases as $L$ decreases ($B_z$
is the axial magnetic field component and $R$ the `minor radius' of the
tube). As the (axial) magnetic flux is conserved we then have for the free
magnetic energy
 \begin{equation}
W_f = \int \frac{B^2_\phi}{2\m} d^3 \vec{r} \propto L^{-1}.    
 \label{1}
\end{equation}

Let us apply the Poynting theorem to a loop anchored in converging motion
at the photospheric boundary. It then follows (from the term with $\vec{v}
\cdot \vec{B} \vec{B} \cdot d \vec{S}$, velocity $\vec{v}$, magnetic field
$\vec{B}$, surface element $d\vec{S}$)  that during a temporal increase in
total magnetic energy the legs of the loop diverge away from each other
into the corona. 

\begin{figure}
\epsscale{1}
\plotone{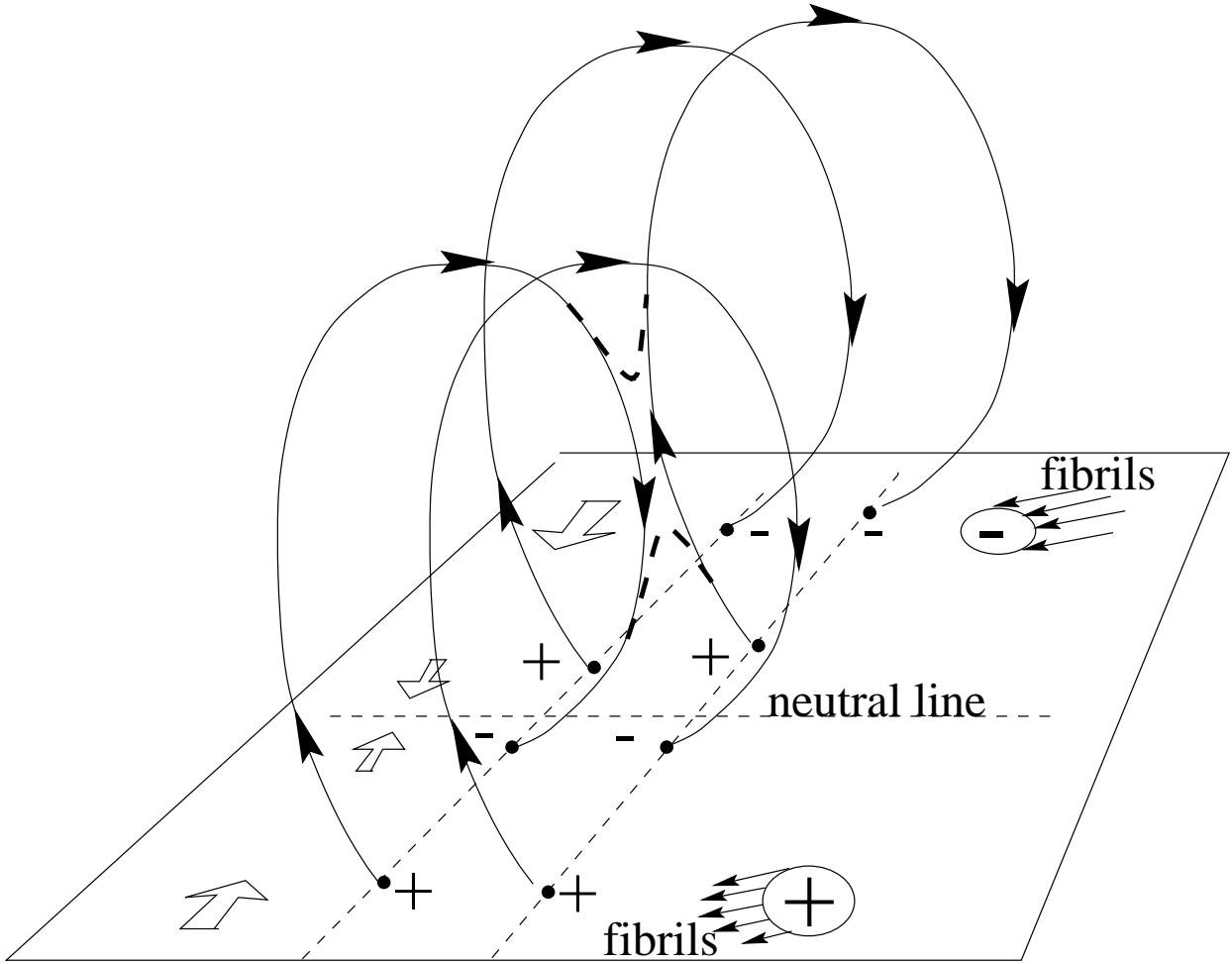}
\caption{Perspective view of four bipolar loops with the same sense of
helicity and migrating towards the neutral
line separating domains of dominant positive (+) and negative
(--) polarity (marked by a small circle from (into) which 
H-$\alpha$ fibrils start (end)). Reconnection is indicated with 
heavy dashed lines. 
\label{f1a}} 
\end{figure}

\begin{figure}
\epsscale{1}
\plotone{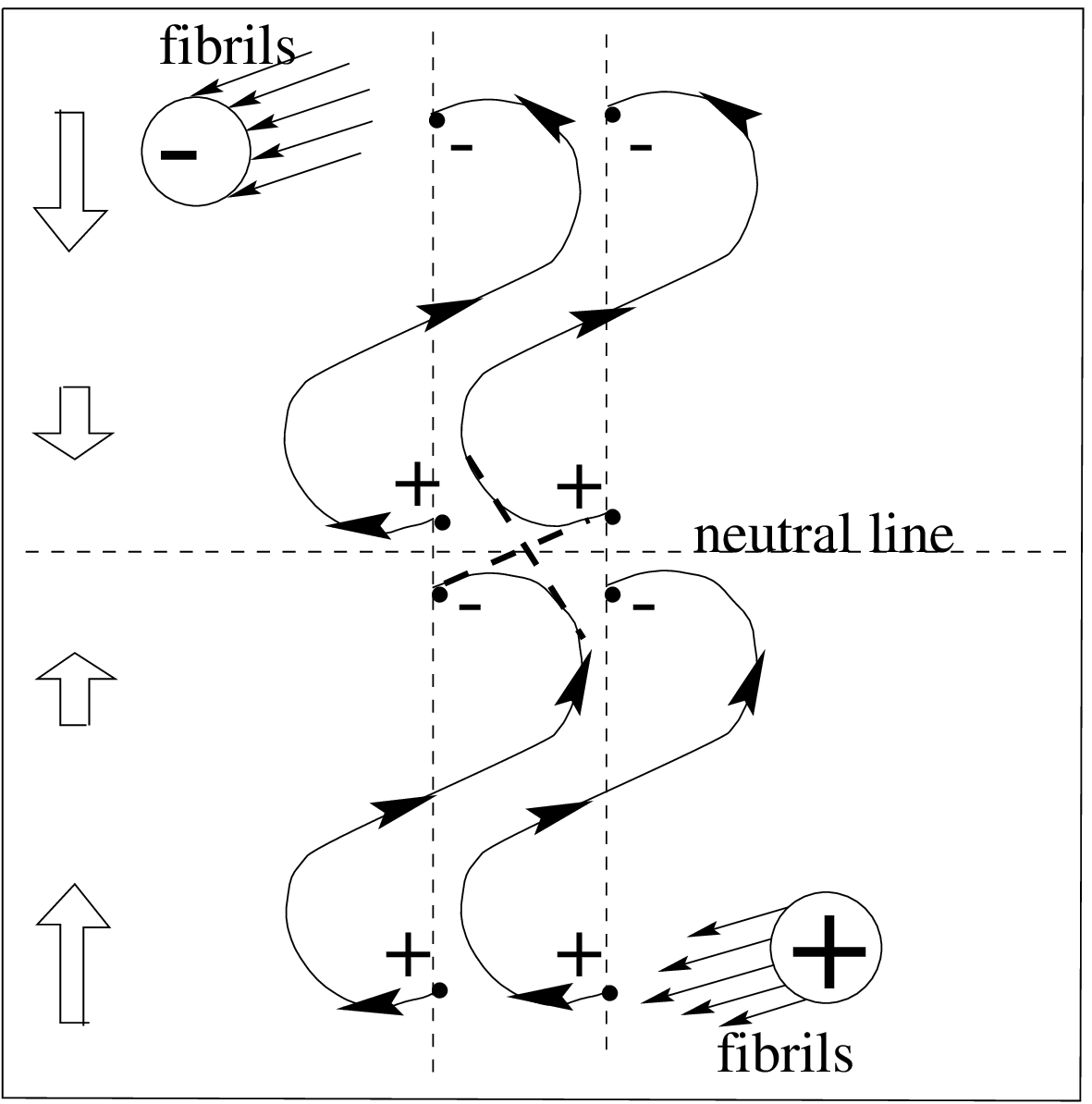}
\caption{Top view of Fig.~\ref{f1a}. Heavy dashed  
curves indicate
reconnection. \label{f1b}} 
\end{figure}

A {\it force free} structure can only store a certain amount of energy
which is dictated by the magnetic field distribution at its bounding
surface, as follows from the scalar magnetic virial theorem
(\cite{aly85,low86}). Asymptotically, an increase in magnetic energy
corresponds to a decreasing tangential component and an opening up of the
field lines. As a result a force free coronal flux tube expands {\it
upward} at the same time when the footpoints converge and the free energy,
defined as the energy stored in coronal electric currents, increases.

Further, as the azimuthal field component in the tube increases with
respect to the axial component the 3-D tube kinks (\cite{fin94}) and
buckles sideways in the same sense as the original twist, thereby
introducing a systematic {\it skew}. The above three effects -- the
diverging legs of an individual loop, its expansion upward, and its
buckling -- are sketched in Figs.~\ref{f1a} and \ref{f1b} where four loops
are anchored in a converging flow pattern such that the internal
separation of each pair of footpoints of the four loops decreases in time.

Actually, the twist and buckling are quantitatively best described by the
concept of magnetic {\it helicity} of a magnetically closed volume
(\cite{bergfiel84,berg84,berg88,moffat}). A flux tube can only be observed
as far as it extends above the photosphere and, strictly speaking, this
part is not bounded by a magnetic surface. It has been shown
(\cite{bergfiel84}) that in this case one can define a relative helicity
which essentially is the coronal part. In ideal MHD the helicity of any
flux tube is conserved under its motion (\cite{wol58}) and this underlies
the twisted shape of the flux tubes in Figs.~\ref{f1a} and \ref{f1b},
where, to be definite, we have assumed a direction of the current on the
magnetic axis of each tube {\it parallel} to the magnetic field (and a
return current on the surface of the tube). 

\subsection{Multiple flux tubes in converging flow}

\begin{figure}
\epsscale{1}
\plotone{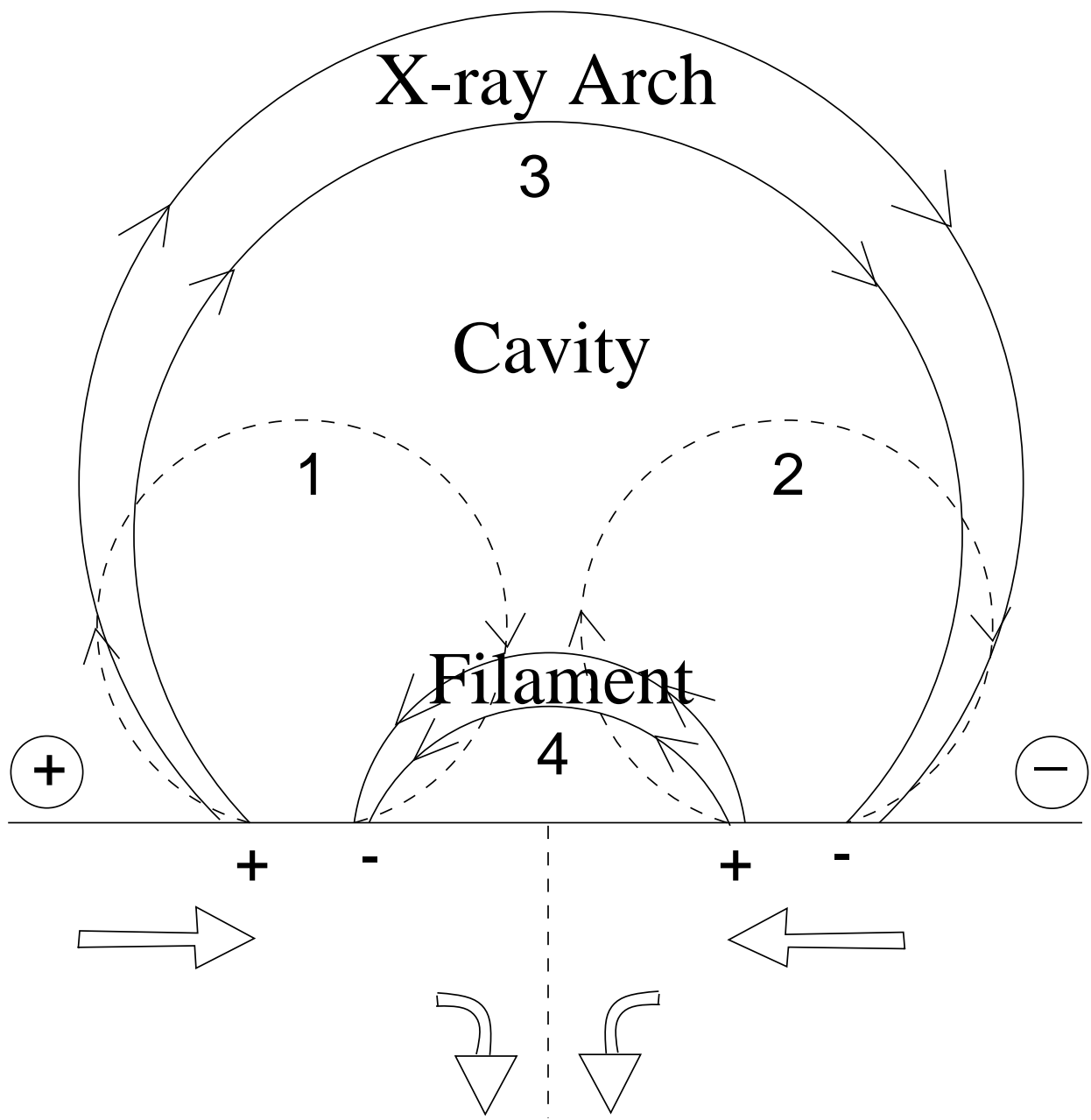}
\caption{Side view of Figs.~\ref{f1a} and \ref{f1b} in the
direction of   the
neutral line, before (dashed) and after (drawn) reconnection.
\label{f2a}} 
\end{figure}

\begin{figure}
\epsscale{1}
\plotone{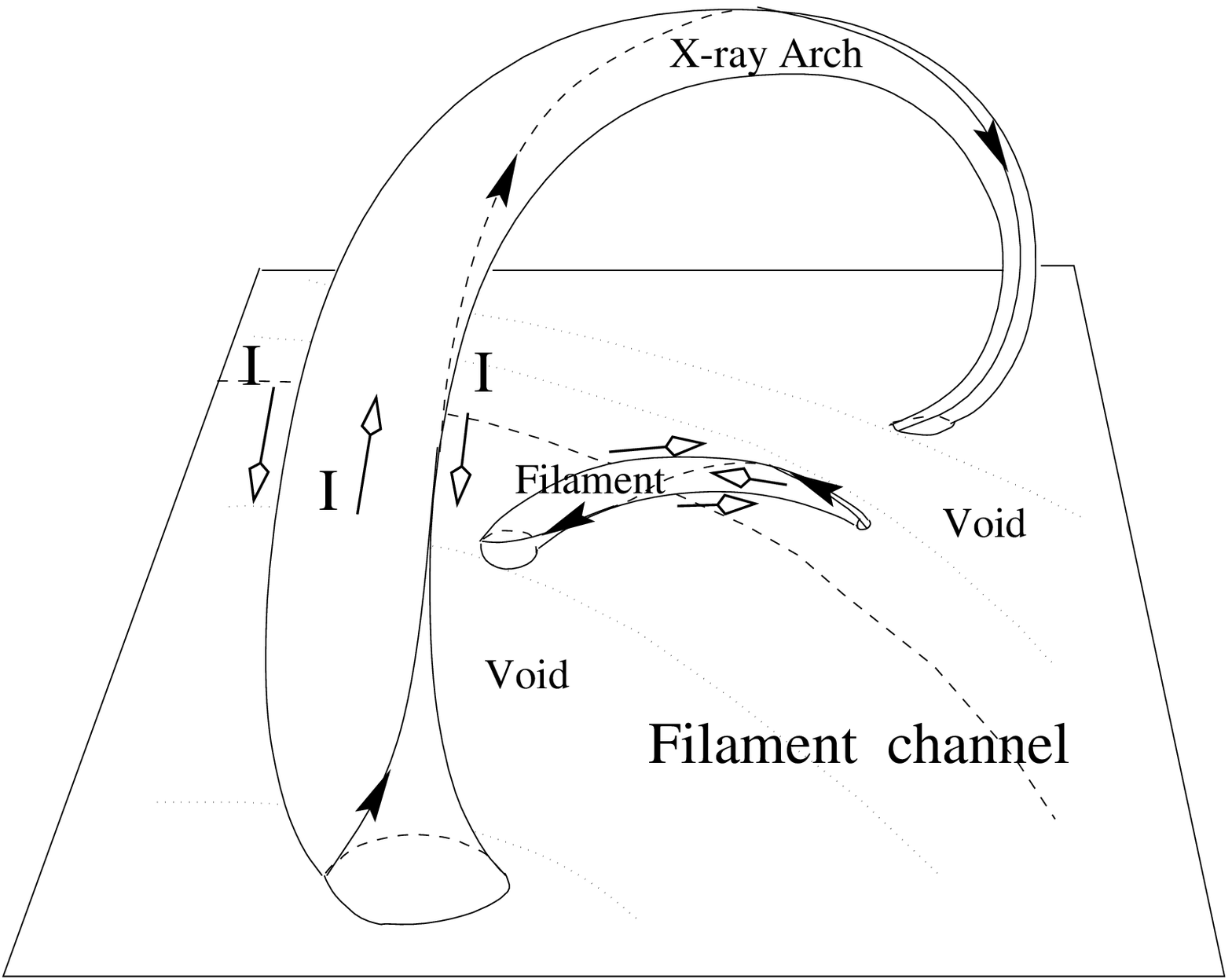}
\caption{Perspective view of a single elementary loop in
the filament and in the overlying arcade. The directions of
the electric currents $I$ are indicated with open arrows. 
\label{f2b}} 
\end{figure}

We now drop the ideal MHD assumption and allow for magnetic reconnection
to take place. In this case one can still define the total magnetic
helicity of the system of reconnecting flux tubes, which, according to
Taylor's conjecture is a conserved quantity (\cite{tay74,tay86}).  In
Figs.~\ref{f1a} and \ref{f1b} a number of flux tubes with parallel
orientation and the same sign of magnetic helicity are anchored in a
converging flow pattern on both sides of the line of convergence -- the
magnetic neutral line between two largely unipolar regions.  As the
helicity in a force free flux tube causes a symmetry breaking of the shape
of each tube as sketched in Figs.~\ref{f1a} and \ref{f1b}, reconnection
between tubes on different sides of the neutral line leads to a system of
inner and outer arches. Figs.~\ref{f2a} and \ref{f2b} picture the relaxed
state after reconnection between two of the loops in Figs.~\ref{f1a} and
\ref{f1b} has occurred. Note that the inner tube is highly sheared
although no shearing motion has been applied. Thus reconnection leads to a
sheared inner row of `neutralised', coaxial current loops. Moreover, the
inner strands have a magnetic field orientation as is observed in
quiescent filaments, that is they are anchored in parasytic polarities. 
Note that eventually the flux of these parasytic polarities is cancelled
as they submerge in the converging flow pattern, and the dominant
polarities on both sides of the neutral line are `cleaned'. In this
picture the elements of a filament are episodically destroyed and
replaced. Also, further reconnection between the inner strands can lead to
a few extended loops in the axial direction, bridging the region of
downflow and escaping destruction.  Finally, the outer row of reconnected
loops have a twist which agrees with the observed streaming direction of
fibrils bordering a filament channel. 

On average the radiative loss in a prominence should balance the heating,
which we assume to derive from magnetic reconnection of the loops in the
converging flow pattern.  Putting the average heating rate per unit volume
equal to the Poynting flux one obtains
 \begin{equation}
E_H \approx \f{v B^2}{2 \m D} =  3.58 \cdot 10^{-4} \mbox{ W
m}^{-3}
     \label{11} \end{equation} for a global convergence speed $v \approx
0.1$ km/s, a typical prominence field of 0.003 T and a horizontal width
transverse to the neutral line of $D\approx 1$ Mm.  Indeed this rate is of
the same order as radiative cooling in the central part of the prominence,
which for an electron density $n_e = 10^{17}$ m$^{-3}$ and a temperature
of 8000 K (the Hvar reference model, \cite{tand95}) amounts to $E_{rad}
\approx 3 \cdot 10^{-4}$ W m$^{-3}$. 

Now, energy release by reconnection occurs in bursts with a typical time
scale of a few Alfv\'en crossing times of the reconnecting structure. Let
us assume that reconnection events occur randomly inside thin flux tubes
crossing the neutral line at a rate of about $10^{-3}$ Hz and that each
takes 1 s (corresponding to a reconnection over 100 km and an average
converging flow of 0.1 km/s). Then the instantaneous heating rate in the
flux tube goes up by a factor $10^3$ and reaches a value 0.3 W m$^{-3}$.
At this rate and for a gas pressure $p= 0.02$ N m$^{-2}$, the quantity
$E_H/p^2$ is nearly an order of magnitude larger than the maximum in the
quantity $\Psi(T)/(2k_BT)^2$ at which steady state coronal loops can exist
(\cite{oord97}, $\Psi(T)$ is the radiative loss function). Clearly, such a
heating is so impulsive that it must lead to strong evaporation of
chromospheric material into the loop and subsequent radiative cooling
below $10^4$ K (cf. also \cite{poland}). The condensed gas becomes visible
as part of the filament, and can be sustained against gravity in a dip of
the twisted flux tube (\cite{amari91,demoul93}), at least temporarily as
long as the tube remains above the chromosphere. 

\section{Coronal cavity and  arcade}

Heating by reconnection in the outer loops would lead to an X-ray arcade. 
From the geometry of the reconnecting loops (Fig.~\ref{f2a})  it follows
that the sites of lower and upper reconnection are separated by a
quiescent volume or cavity. Reconnection in the overlying arcade again
causes heating at an average rate given by Eq.~(\ref{11}). However as both
the field strength is lower and the dimensions are larger (typically
$B=10^{-4}$ T and $D = 30$ Mm)  the heating rate per unit volume is at
least four orders of magnitude smaller, $E_H \approx 1.3 \cdot 10^{-8}$ W
m$^{-3}$.  For such values steady hot coronal loops can exist without a
problem (\cite{oord97}). Multiple reconnection events then again lead to
evaporation but now a hot coronal arcade is maintained as the footpoints
converge.

Both the filament and the arcade can go unstable because both are
essentially force free equilibrium structures which tend to expand as the
total current increases. The first would show up as a filament flare, the
latter as a Coronal Mass Ejection (CME). Note that the proposed mechanism
would explain why a CME has already an enhanced density before the
explosion occurs.  Also the finite gas pressure may have an important
effect on the CME instability (\cite{zwin87}).

The only practical complication in describing the filament flare is the
large number of neutralised currents, which can not be modelled by a
simple circuit. The occurrence of instability is determined by the
magnitude of the current system which depends on the number of reconnected
bipolar loops, their initial currents, and the footpoint displacement.
Interestingly, in our model each of these quantities would be correllated
in the arcade and in the prominence. This effect may dissolve the
long-standing debate about the relation between CMEs and flares. In the
present model both are indeed `signatures of the same disease'
(\cite{har95}).  Here it is of interest to mention the numerical
simulation of a converging, quadrupolar, arcade (however, without
submerging central flow) presented by Dr. Uchida at the recent IAU Symp.
No. 188 in Kyoto (\cite{uchida}).
   
\section{Discussion}

Quiescent prominences in the same hemisphere are observed to have the same
handedness independent of solar cycle. Here we have shown how a converging
flow in an ordered series of bipoles with the same sense of helicity would
lead to the formation of prominences with the observed inverse polarity
and the same handedness (dextral/sinistral). Magnetic loops rising upward
inside the sun experience a Coriolis force and develop the same sense of
helical distortion in the same hemisphere.  If these distortions lead to a
magnetic helicity the sense of the magnetic helicity would be independent
of the loop orientation and, consequently, independent of the solar cycle.

Both the quasi-steady high density at coronal temperatures in an arcade
and the non-steady high density at a low temperature in a filament are
caused by reconnections, the difference in appearance being the result of
a dramatic difference in volume heating rates caused by a difference in
dimensions. The observed upward flows (5 km/s in CIV and 0.5 km/s in
H-$\alpha$, \cite{schmie}) are entirely consistent with cooling gas flows
after evaporation from impulsive reconnection events. Also the excellent
fit of a prominence shape by a linear force free field (\cite{schmie89})
is consistent with magnetic relaxation satisfying Taylor's conjecture.
Finally, in our model the horizontal component of an extended
non-potential field -- which is held to cause alignment of fibrils at the
border of a filament channel (\cite{gaiz}) -- derives from electric
currents in the outer reconnected loops while reconnections of sheared
inner loops lead to the appearance of a filament. 

The present model starts off from a collection of dipoles without any net
magnetic flux in the direction of the neutral line. As soon as chiral
symmetry is lost such an `axial' component is created in the corona (see
Figs.~\ref{f1a} and \ref{f1b}), and persists after reconnection. However,
the {\it net} axial flux, summed over the arch and the filament, remains
zero. This seems to be in conflict with observations at lower latitudes
(\cite{martal}) but in agreement with polar crown arcades (\cite{mcal}).

How does our proposal relate to existing models? Ours is essentially
dynamical and depends on ongoing reconnection between bipolar loops of the
same helicity in a converging flow pattern. It therefore differs from the
model by van Ballegooijen \& Martens (1989) which requires a shearing
motion. The quasistatic support of gas against gravity in our model is in
dips as in the Kippenhahn-Schl\"uter (KS) model (\cite{kip}). A difference
with the latter is that the existence of (largely force free) currents is
essential to the present mechanism while force free currents are absent in
the KS case.  Therefore the dips can be primarily of a force free nature.
This practically force free aspect is an essential ingredient also of the
Kuperus-Raadu (KR) model (\cite{kup}). A difference with the latter model
is that the currents in the present model are largely coronal with coaxial
return currents, while in the KR model the coronal current has a unique
direction and the return current is in the photosphere/chromosphere.

\acknowledgments I would like to thank Sara Martin,  Bert van den Oord, 
Max Kuperus, and the referee for their comments.

\end{document}